\documentclass[pdflatex,sn-basic,Numbered]{sn-jnl}

\usepackage{graphicx}%
\usepackage{multirow}%
\usepackage{amsmath,amssymb,amsfonts}%
\usepackage{amsthm}%
\usepackage{mathrsfs}%
\let\mathcal\mathscr
\usepackage[title]{appendix}%
\usepackage{xcolor}%
\usepackage{textcomp}%
\usepackage{manyfoot}%
\usepackage{booktabs}%
\usepackage{array}
\usepackage{algorithm}%
\usepackage{algorithmicx}%
\usepackage{algpseudocode}%
\usepackage{listings}%
\usepackage{float}
\usepackage[utf8]{inputenc}
\usepackage[T1]{fontenc}
\usepackage{lmodern}
\theoremstyle{thmstyleone}%
\theoremstyle{thmstyletwo}%

\theoremstyle{thmstylethree}%

\begin{document}

\title[Article Title]{Adsorption of Water on Pristine Graphene: A van der Waals Density Functional Study with the vdW-C09 Approach}


\author[1]{\fnm{Aline Oliveira Santos} \sur{Author}}\email{alineos@ufba.br}

\author[2]{\fnm{Bruno H. S. Mendonça} \sur{Author}}\email{brunnohennrique13@gmail.com}
\equalcont{These authors contributed equally to this work.}

\author*[1]{\fnm{Elizane E. de Moraes} \sur{Author}}\email{elizane.fisica@gmail.com}
\equalcont{These authors contributed equally to this work.}

\affil[1]{\orgdiv{Instituto de F{\'i}sica}, \orgname{Universidade Federal da Bahia}, \orgaddress{\street{Campus Universit{\'a}rio de Ondina}, \city{Salvador}, \postcode{40210-340}, \state{Bahia}, \country{Salvador}}}

\affil[2]{\orgdiv{Departamento de F{\'i}sica, ICEX}, \orgname{ Universidade Federal de Minas Gerais}, \orgaddress{\street{Av. Pres. Antônio Carlos, 6627}, \city{Belo Horizonte}, \postcode{31270-901}, \state{Minas Gerais}, \country{Brazil}}}


\abstract{
\textbf{Context} Understanding how water interacts with graphene at the molecular level is essential for advancing nanomaterial applications in filtration, catalysis, and environmental technologies. This study establishes a quantitative baseline for assessing how structural defects, dopants, or surface functionalization may enhance water adsorption, providing insights for the rational design of graphene-based materials in water purification, sensing, and nanofluidic applications.

\textbf{Methods} In this work, we employed density functional theory (DFT) with the vdW-C09 functional to investigate the adsorption of a single water molecule on pristine graphene, accurately accounting for long-range dispersion forces. Three high-symmetry adsorption sites—the center of the hexagonal ring, the C–C bond, and the top site—were explored in combination with three molecular orientations: Down, H-bond, and Up configurations. The calculated adsorption energies range from –93 to –145 meV (milli–electron volts), indicating that the interaction is dominated by weak van der Waals forces characteristic of physisorption. The most stable configuration corresponds to the Down orientation above the center of the hexagonal ring, with an adsorption energy of $-145$ meV and an equilibrium distance of 3.27 \AA (ångström), defined as the vertical separation between the oxygen atom of the water molecule and the graphene plane. These results are in close agreement with previous theoretical studies and confirm the non-reactive and hydrophobic nature of pristine graphene.
}




\maketitle

\section{Introduction}\label{sec1}

Due to its remarkable electronic, mechanical, and surface area properties, graphene has become the foundation of modern nanotechnology \cite{catania2021review,ganguly2024graphene,schmaltz2024graphene,xu2024graphene,guha2025graphene,kanti2025graphene,sahoo2024recent}. In particular, its application in aqueous systems, such as desalination membranes, advanced nanofilters, and environmental catalysis devices, demonstrates revolutionary potential in the field of water treatment \cite{zubair2024challenges,tiwary2024graphene,perreault2015environmental,hu2022construction,neeraja2025comprehensive,zeng2021engineered}. However, the performance and efficiency of these nanodevices are intrinsically dictated by the fundamental nature of the interface between water and graphene. The precise interaction of the water molecule with the inert, nonpolar graphene surface is a key factor in controlling wettability, nanoscale fluid transport, and the mechanism of contaminant adsorption \cite{ahmed2019challenges,moraes2017transport,hodges2018challenges,werber2016materials,fatima2022tunable,zubair2024review,garcia2022interfacial,@10.1021/acs.jpcb.3c02889,10.1021/acs.jpcc.4c00078}.

Although important, accurately describing the interaction between water and graphene remains a theoretical challenge \cite{de2019density,han2021review,dong2018interface}. The predominant interactions are weak, non-covalent dispersion forces, commonly referred to as van der Waals forces. Traditional Density Functional Theory (DFT) methods often do not adequately represent these forces \cite{nezval2023dft}. The lack of a thorough understanding of the preferred adsorption sites and molecular orientations of water on the graphene surface limits the rational optimization of carbonaceous materials. To provide a reliable microscopic image, it is crucial to utilize methodologies that accurately capture these long-range scattering effects \cite{dong2018interface,garcia2022interfacial,verma2024experimental,gbadamasi2021interface,bjorneholm2016water,sacchi2023water,li2015two,kang2024graphene}.

Accurate determination of the interaction energy between water molecules and graphene, resulting from the subtle balance between water-surface adhesion forces and intermolecular hydrogen bonds, is essential for a realistic description of water behavior in graphene membranes \cite{hamada2012adsorption,peng2017review,ma2011adsorption,voloshina2011physisorption,melios2018water,sacchi2023water,wehling2008first,leenaerts2009water}. The water-graphene interaction is characteristically weak, yet it plays a decisive role in modulating the structural and dynamic properties of water confined within carbon nanostructures \cite{holt2006fast,hummer2002water}. The main mechanism involved is the van der Waals (vdW) interaction, which governs the adsorption of water molecules on the graphene surface \cite{li2017molecular}. Additionally, electrostatic and polarization effects are relevant: the presence of adsorbed water molecules induces significant polarization in the carbon network, contributing to the overall attractive potential between water and graphene \cite{brandenburg2019interaction,ahlstrom1989molecular,lamoureux2006polarizable}. The weak nature of the adsorption is directly reflected in the electronic structure of graphene \cite{hamada2012adsorption}. The presence of water molecules does not induce appreciable doping, and the system's electronic bands remain practically unchanged compared to those of intrinsic graphene. Only a slight splitting of bands is observed, associated with symmetry breaking and the introduction of additional energy levels from the molecular orbitals of water \cite{cao2010transition,vanin2010first,michaelides2006density,valle2024accuracy}. 

The structural organization of water on surfaces or under confinement is governed by a delicate balance between adsorption interaction on the surface and intermolecular hydrogen bonds \cite{holt2006fast,falk2010molecular,singh2018critical,shuvo2025hydrodynamic,xu2025advanced}. Thus, the rigorous quantification of these interactions is a fundamental requirement for the accurate modeling of water adsorption on graphene, providing the necessary theoretical basis for understanding the ultrafast (or anomalous) flow of water in graphene membranes, which is the phenomenon that underpins the potential of these structures in advanced desalination and water purification processes \cite{cohen2012water,surwade2015water,homaeigohar2017graphene,konatham2013simulation,aghigh2015recent}.

Although the adsorption of water on graphene has been addressed in previous theoretical works, reported binding energies and equilibrium distances still vary depending on the treatment of dispersion interactions. In this study, we employ the vdW-C09 functional, which has been shown to provide an improved description of long-range van der Waals forces compared to standard GGA approaches, often reducing overbinding and distance inaccuracies. Our objective is not to introduce a new methodology, but to establish a consistent and well-characterized reference for water adsorption on pristine graphene, which can serve as a reliable baseline for future investigations of defected, doped, or functionalized graphene surfaces.

In this work, we therefore provide a systematic description of the adsorption of a single water molecule on a perfect graphene sheet using DFT with the vdW-C09 functional. We investigate three high-symmetry adsorption sites—the center of the hexagonal ring, the C–C bond, and the top site—and three distinct molecular orientations: down, H-bond, and up. By comparing adsorption energies and equilibrium distances, this study identifies the thermodynamically most favorable configurations and structural characteristics of water in contact with graphene. The results establish a validated reference dataset that contributes to a fundamental understanding of how nanoscale surface properties influence water behavior, providing a solid theoretical basis for the rational design of graphene-based materials for filtration, catalysis, and environmental nanotechnology.

\section{Computational details}\label{sec2}

Our calculations were based on the density functional theory 
(DFT)  \cite{PhysRev.140.A1133} as implemented in the SIESTA code version 4.1-b4 \cite{soler2002siesta, sanchez1997density}. We employed the well-known exchange–correlation functional of vDW-C09 \cite{cooper2010van}. Spin polarization was included in all calculations. We used the norm-conserving pseudopotentials in the Kleinman-Bylander factorized form \cite{testeKB,martins91} with a double-zeta plus polarization (DZP) basis set for C, H, and O atoms. The real-space integration grid was defined by an energy cutoff of 350 Ry. Structural optimizations were carried out until the residual forces on all atoms were smaller than 0.05 eV/\AA$^{-1} $. Self-consistent field (SCF) calculations were considered converged when the total energy variation between successive iterations was below $10^ {-5}$ eV, and the density-matrix variation was below $10^{-5}$. 

Although the water–graphene system is formally a closed-shell system, spin polarization was included in all calculations to avoid artificial constraints and to allow for possible symmetry-breaking or subtle electronic rearrangements upon adsorption. In all cases, the final converged solutions corresponded to a non-magnetic ground state with zero net spin polarization.

A graphene supercell containing 144 carbon atoms was employed with periodic boundary conditions, and a vacuum region of approximately 15 \AA was introduced along the out-of-plane direction to prevent spurious interactions between periodic images

For the interaction energy calculations, we used the following equation through BSSE (basis set superposition error) corrected for all calculations with the counterpoise method :

\begin{equation}
    E=E(A+B)-E(A+B_{ghost}) -E(B+A_{ghost})
\end{equation}

This correction is performed starting from the initial geometry of the AB system and calculating the total energy of system A, considering the whole set of base functions, where the set of base functions B is in the position corresponding to system B, without the explicit presence of the atoms. The same occurs in the calculation of system B. The system with negative binding energies implies an attractive interaction.

\section{Results and discussion}\label{sec3}

\begin{figure}[htbp]
    \centering
    \includegraphics[width=\linewidth]{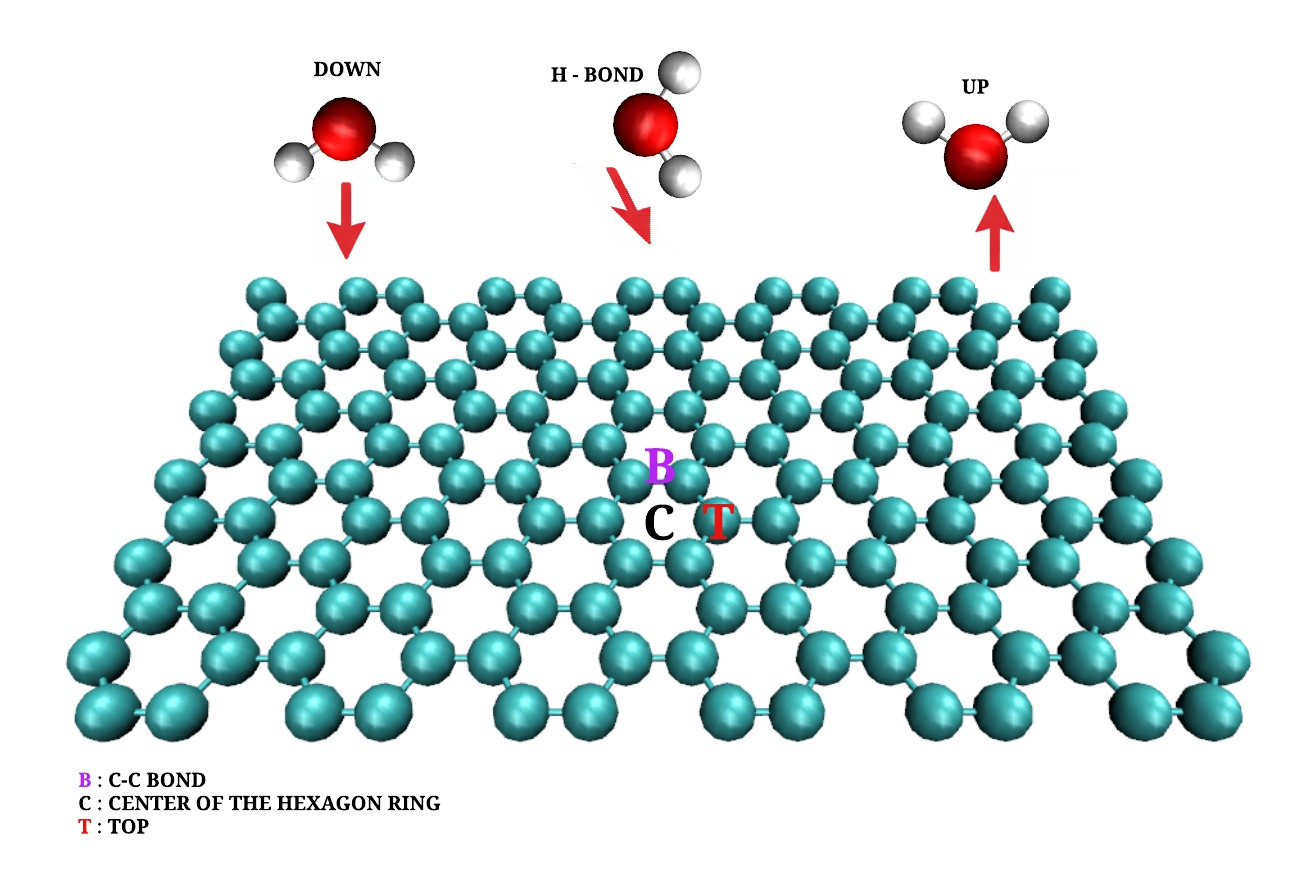}
    \caption{ Optimized geometries of a single water molecule adsorbed on pristine graphene obtained from DFT calculations using the VDW-C09 functional. Three adsorption sites were considered: (a) center of the hexagonal ring, (b) C–C bond, and (c) top of a carbon atom. Oxygen atoms are shown in red, hydrogen in white, and carbon in blue.}
    \label{Fig1}
\end{figure}

The interaction between a single water molecule and pristine graphene was investigated using DFT calculations within the VDW-C09 functional, which properly accounts for long-range dispersion forces. Three adsorption sites were considered: the center of the hexagonal ring, the C–C bond, and the top site, along with three molecular orientations: Down, H-bond, and Up, as seen in Figure \ref{Fig1}. The corresponding adsorption energies (E$_{ads}$) and equilibrium distances between the oxygen atom of the water molecule and the graphene surface are summarized in Table \ref{table}. The equilibrium distance is defined as the vertical distance from the oxygen atom of a water molecule to the graphene plane, rather than the shortest O–C interatomic distance.

\begin{table}[htbp]
\centering
\caption{Adsorption energies ($E_{\text{ads}}$) and equilibrium distances ($d$) of a single water molecule on pristine graphene for different adsorption sites and molecular orientations, obtained from DFT calculations with the VDW-C09 functional.}
\label{tab:adsorption}
\begin{tabular}{lccccccc}
\hline
\textbf{Orientation} & \multicolumn{2}{c}{\textbf{Center of Hexagonal Ring}} & \multicolumn{2}{c}{\textbf{C–C Bond}} & \multicolumn{2}{c}{\textbf{Top}} \\

 & $E_{\text{ads}}$ (meV) & $d$ (\AA) & $E_{\text{ads}}$ (meV) & $d$ (\AA) & $E_{\text{ads}}$ (meV) & $d$ (\AA) \\
\hline
Down    & -145 & 3.27 &-141 & 2.80 & -137 & 2.90 \\
H-bond  & -128 & 2.90 & -133 & 2.60 & -134 & 2.70 \\
Up      & -98  & 3.07 & -94  & 3.10 & -93  & 3.30 \\
\hline
\end{tabular}
\label{table}
\end{table}

Our calculations reveal that the interaction between water and pristine graphene is weak, with adsorption energies ranging from –93 to $-145$ meV. Such values are characteristic of physisorption, confirming that water molecules adhere to graphene mainly through van der Waals dispersion forces. This finding aligns with previous reports and supports the hydrophobic character typically observed for pristine graphene surfaces.

Among the three orientations considered, the Down configuration—where the hydrogen atoms of the water molecule point toward the graphene surface—was found to be the most stable, particularly when the molecule is placed over the center of the hexagonal ring (–145 meV). This preference can be understood in terms of the molecular dipole of water interacting with the polarizable $\pi$-electron cloud of graphene. In the Down orientation, the hydrogen atoms, corresponding to the positive end of the dipole, are positioned closer to the surface, enhancing dipole–induced dipole and polarization contributions in addition to dispersion forces.

The H-bond configuration exhibits intermediate stability, with adsorption energies around –130 meV and slightly shorter equilibrium distances (down to 2.6 \AA), consistent with weak local polarization effects but not indicative of chemical bonding. In contrast, the Up configuration, in which the oxygen atom faces the surface, is the least stable (–93 to –98 meV) and corresponds to the largest separations between the molecule and the graphene plane, reflecting a less favorable electrostatic alignment of the molecular dipole with the surface.

The equilibrium distances obtained (2.6–3.3 \AA), together with the weak adsorption energies, are characteristic of physisorption dominated by van der Waals interactions. Moreover, the small energy differences among adsorption sites indicate that pristine graphene provides a nearly homogeneous, nonpolar surface with no strongly preferred binding positions. In this weak-binding regime, the interaction is nonreactive, with any electronic rearrangement limited to minor polarization effects rather than significant charge transfer or covalent bonding.

The adsorption energies obtained in this work are in excellent agreement with previous DFT studies that also describe water–graphene interactions as weakly bound systems governed by van der Waals forces \cite{leenaerts2009water,ma2011adsorption,raj2022effect,sacchi2023water,hamada2012adsorption}. Some works \cite{leenaerts2009water,ma2011adsorption} reported adsorption energies ranging from –100 to –150 meV and equilibrium distances around 3.0\AA using vdW-corrected functionals, consistent with our results. Similarly, works  \cite{raj2022effect,sacchi2023water,hamada2012adsorption}, found comparable values using PBE-D2 and optB88-vdW approaches, further confirming that the Down configuration—where the hydrogen atoms of the water molecule face the graphene surface—is the most stable orientation. In addition, experimental observations also support this weak physisorption behavior. Contact angle measurements and surface wetting studies on pristine graphene consistently indicate hydrophobic behavior, with contact angles typically above 85, implying only minor adhesion of water to the surface \cite{carlson2024modeling,pizzochero2024one,temmen2014hydration}. These findings reinforce that the water–graphene interaction is dominated by dispersion forces and that chemical bonding does not occur under ambient conditions.

Finally, the small differences among adsorption sites reflect the uniform, nonpolar nature of the graphene lattice. The optimized geometries confirm that the water molecule remains nearly intact after adsorption, with no evidence of surface structural distortion. In this context, the interaction is best described as physisorption, with negligible electronic rearrangement beyond weak polarization. These results also indicate that surface modifications—such as the introduction of structural defects, heteroatom dopants, or oxygen-containing functional groups—could substantially alter the adsorption landscape, increasing interaction strength and potentially enhancing the hydrophilicity of graphene-based materials. Such tunability is particularly relevant for applications in water purification, sensing, and catalysis.

\section{Conclusions}\label{sec4}

In this study, we investigated how a single water molecule interacts with pristine graphene using DFT calculations with the VDW-C09 functional. The results indicate that the interaction is weak, primarily governed by van der Waals forces, with adsorption energies ranging from –93 to –145 meV. Among all configurations analyzed, the most stable arrangement corresponds to the Down orientation of the water molecule positioned above the center of the hexagonal ring, with an adsorption energy of –145 meV and an equilibrium distance of approximately 3.27 \AA. In this geometry, the water molecule is oriented toward the graphene surface, which favors slightly stronger dispersion interactions. The H-bond and Up configurations are less stable, presenting smaller binding energies and longer molecule–surface separations. The calculated equilibrium distances, between 2.6 and 3.3 \AA, confirm that the molecule remains physically adsorbed, without forming chemical bonds or significantly altering the graphene surface. The small variation in adsorption energy across different sites also reflects the homogeneous and non-reactive nature of pristine graphene, consistent with its experimentally observed hydrophobic behavior. Overall, these results provide a clear and intuitive understanding of how water interacts with an ideal graphene sheet. Although the interaction is weak, it serves as a foundation for exploring how defects, dopants, or functional groups could be introduced to tailor graphene’s surface chemistry. Such modifications can enhance water adsorption, opening new opportunities for applications in water purification, catalysis, and nanoscale fluid control.

\backmatter

\section{Acknowledgements}
This work is funded by the Brazilian scientific agencies Fundação de Amparo à Pesquisa do Estado da Bahia (FAPESB), Fundação de Amparo à Pesquisa do Estado de Minas Gerais (FAPEMIG), Comissão de Aperfeiçoamento do Pessoal de Ensino Superior (CAPES), Conselho Nacional de Desenvolvimento Científico e Tecnológico (CNPq) and the Brazilian Institute of Science and Technology (INCT) in Carbon Nanomaterials, with collaboration and computational support from Universidade Federal da Bahia (UFBA). In addition, the authors acknowledge the National Laboratory for Scientific Computing (LNCC/MCTI, Brazil) for providing HPC resources on the SDumont supercomputer, which contributed to the research results reported in this paper. URL: http://sdumont.lncc.br. Finally,  EEM appreciates Edital PRPPG 010/2024 Programa de Apoio a Jovens Professores(as)/Pesquisadores(as) Doutores(as) - JOVEMPESQ Project 24460.

\section*{Declarations}

Not applicable


\begin{thebibliography}{68}
\providecommand{\natexlab}[1]{#1}
\providecommand{\url}[1]{{#1}}
\providecommand{\urlprefix}{URL }
\providecommand{\doi}[1]{\url{https://doi.org/#1}}
\providecommand{\eprint}[2][]{\url{#2}}
 \bibcommenthead

\bibitem[{Catania, Federica and Marras, Elena and Giorcelli, Mauro and Jagdale, Pravin and Lavagna, Luca and Tagliaferro, Alberto and Bartoli, Mattia(2021)Catania, Marras, Giorcelli, Jagdale, Lavagna, Tagliaferro, and Bartoli}]{catania2021review}
Catania F, Marras E, Giorcelli M, Jagdale P, Lavagna L, Tagliaferro A, and Bartoli M (2021) A review on recent advancements of graphene and graphene-related materials in biological applications. Applied Sciences 11(2):614. \doi{https://doi.org/10.3390/app11020614}

\bibitem[{Ganguly, Sharmi and Sengupta, Joydip(2024)Ganguly and Sengupta}]{ganguly2024graphene}
Ganguly S and Sengupta J (2024) Graphene-based nanotechnology in the internet of things: a mini review. Discover Nano 19(1):110. \doi{https://doi.org/10.1186/s11671-024-04054-0}

\bibitem[{Schmaltz, Thomas and Wormer, Lorenzo and Schmoch, Ulrich and D{\"o}scher, Henning(2024)Schmaltz, Wormer, Schmoch, and D{\"o}scher}]{schmaltz2024graphene}
Schmaltz T, Wormer L, Schmoch U, and D{\"o}scher H (2024) Graphene roadmap briefs (no. 3): meta-market analysis 2023. 2D Materials 11(2):022002. \doi{10.1088/2053-1583/ad1e78}

\bibitem[{Xu, Yang and Liu, Enke(2024)Xu and Liu}]{xu2024graphene}
Xu Y and Liu E (2024) Graphene: Two decades of revolutionizing material science. \doi{https://doi.org/10.59717/j.xinn-mater.2024.100059}

\bibitem[{Guha, Spandan and Chakrabarty, Shanta(2025)Guha and Chakrabarty}]{guha2025graphene}
Guha S and Chakrabarty S (2025) Graphene and its derivatives (go, rgo and gqd): a comprehensive review of their role in combating covid-19. Advances in Physics: X 10(1):2435278. \doi{https://doi.org/10.1080/23746149.2024.2435278}

\bibitem[{Kanti, Praveen Kumar and HG, Prashantha Kumar and Wanatasanappan, V Vicki and Kumar, Abhinav and Regasa, Melkamu Biyana(2025)Kanti, HG, Wanatasanappan, Kumar, and Regasa}]{kanti2025graphene}
Kanti PK, HG PK, Wanatasanappan VV, Kumar A, and Regasa MB (2025) Graphene's frontier in aerospace: current applications, challenges, and future directions for space engineering. Nanoscale Advances 7(12):3603--3618. \doi{10.1039/D4NA00934G}

\bibitem[{Sahoo, Prasanta Kumar and Kumar, Niraj and Jena, Anirudha and Mishra, Sujata and Lee, Chuan-Pei and Lee, Seul-Yi and Park, Soo-Jin(2024)Sahoo, Kumar, Jena, Mishra, Lee, Lee, and Park}]{sahoo2024recent}
Sahoo PK, Kumar N, Jena A, Mishra S, Lee CP, Lee SY, and Park SJ (2024) Recent progress in graphene and its derived hybrid materials for high-performance supercapacitor electrode applications. RSC advances 14(2):1284--1303. \doi{10.1039/D3RA06904D}

\bibitem[{Zubair, Muhammad and Roopesh, MS and Ullah, Aman(2024)Zubair, Roopesh, and Ullah}]{zubair2024challenges}
Zubair M, Roopesh M, and Ullah A (2024) Challenges and prospects: graphene oxide-based materials for water remediation including metal ions and organic pollutants. Environmental Science: Nano 11(9):3693--3720. \doi{https://doi.org/10.1039/D4EN00143E}

\bibitem[{Tiwary, Saurabh Kr and Singh, Maninderjeet and Chavan, Shubham Vasant and Karim, Alamgir(2024)Tiwary, Singh, Chavan, and Karim}]{tiwary2024graphene}
Tiwary SK, Singh M, Chavan SV, and Karim A (2024) Graphene oxide-based membranes for water desalination and purification. npj 2D Materials and Applications 8(1):27. \doi{https://doi.org/10.1038/s41699-024-00462-z}

\bibitem[{Perreault, Fran{\c{c}}ois and De Faria, Andreia Fonseca and Elimelech, Menachem(2015)Perreault, De~Faria, and Elimelech}]{perreault2015environmental}
Perreault F, De~Faria AF, and Elimelech M (2015) Environmental applications of graphene-based nanomaterials. Chemical Society Reviews 44(16):5861--5896. \doi{https://doi.org/10.1039/C5CS00021A}

\bibitem[{Hu, Huawen and Wen, Wu and Ou, Jian Zhen(2022)Hu, Wen, and Ou}]{hu2022construction}
Hu H, Wen W, and Ou JZ (2022) Construction of adsorbents with graphene and its derivatives for wastewater treatment: a review. Environmental Science: Nano 9(9):3226--3276. \doi{https://doi.org/10.1039/D2EN00248E}

\bibitem[{Neeraja, SM and Bindhu, B(2025)Neeraja and Bindhu}]{neeraja2025comprehensive}
Neeraja S and Bindhu B (2025) A comprehensive review on two-dimensional materials for water treatment applications. ES Energy and Environment 28:1411. \doi{10.30919/ee1411}

\bibitem[{Zeng, Minxiang and Chen, Mingfeng and Huang, Dali and Lei, Shijun and Zhang, Xuan and Wang, Ling and Cheng, Zhengdong(2021)Zeng, Chen, Huang, Lei, Zhang, Wang, and Cheng}]{zeng2021engineered}
Zeng M, Chen M, Huang D, Lei S, Zhang X, Wang L, and Cheng Z (2021) Engineered two-dimensional nanomaterials: an emerging paradigm for water purification and monitoring. Materials Horizons 8(3):758--802. \doi{https://doi.org/10.1039/D0MH01358G}

\bibitem[{Ahmed, M and Giwa, A and Hasan, SW(2019)Ahmed, Giwa, and Hasan}]{ahmed2019challenges}
Ahmed M, Giwa A, and Hasan S (2019) Challenges and opportunities of graphene-based materials in current desalination and water purification technologies. Nanoscale Materials in Water Purification pp 735--758. \doi{https://doi.org/10.1016/B978-0-12-813926-4.00033-1}

\bibitem[{Moraes, Elizane E and Coutinho-Filho, Maur{\'\i}cio D and Batista, Ronaldo JC(2017)Moraes, Coutinho-Filho, and Batista}]{moraes2017transport}
Moraes EE, Coutinho-Filho MD, and Batista RJ (2017) Transport properties of hydrogenated cubic boron nitride nanofilms with gold electrodes from density functional theory. ACS omega 2(4):1696--1701. \doi{10.1021/acsomega.7b00061}

\bibitem[{Hodges, Brenna C and Cates, Ezra L and Kim, Jae-Hong(2018)Hodges, Cates, and Kim}]{hodges2018challenges}
Hodges BC, Cates EL, and Kim JH (2018) Challenges and prospects of advanced oxidation water treatment processes using catalytic nanomaterials. Nature Nanotechnology 13(8):642--650. \doi{https://doi.org/10.1038/s41565-018-0216-x}

\bibitem[{Werber, Jay R and Osuji, Chinedum O and Elimelech, Menachem(2016)Werber, Osuji, and Elimelech}]{werber2016materials}
Werber JR, Osuji CO, and Elimelech M (2016) Materials for next-generation desalination and water purification membranes. Nature Reviews Materials 1(5):1--15. \doi{https://doi.org/10.1038/natrevmats.2016.18}

\bibitem[{Fatima, Jawaria and Shah, Adnan Noor and Tahir, Muhammad Bilal and Mehmood, Tariq and Shah, Anis Ali and Tanveer, Mohsin and Nazir, Ruqia and Jan, Basit Latief and Alansi, Saleh(2022)Fatima, Shah, Tahir, Mehmood, Shah, Tanveer, Nazir, Jan, and Alansi}]{fatima2022tunable}
Fatima J, Shah AN, Tahir MB, Mehmood T, Shah AA, Tanveer M, Nazir R, Jan BL, and Alansi S (2022) Tunable 2d nanomaterials; their key roles and mechanisms in water purification and monitoring. Frontiers in Environmental Science 10:766743. \doi{https://doi.org/10.3389/fenvs.2022.766743}

\bibitem[{Zubair, Muhammad and Farooq, Sadia and Hussain, Ajaz and Riaz, Sadia and Ullah, Aman(2024)Zubair, Farooq, Hussain, Riaz, and Ullah}]{zubair2024review}
Zubair M, Farooq S, Hussain A, Riaz S, and Ullah A (2024) A review of current developments in graphene oxide--polysulfone derived membranes for water remediation. Environmental Science: Advances 3(7):983--1003. \doi{10.1039/D4VA00058G}

\bibitem[{Garcia, Ricardo(2022)Garcia}]{garcia2022interfacial}
Garcia R (2022) Interfacial liquid water on graphite, graphene, and 2d materials. ACS Nano 17(1):51--69. \doi{https://doi.org/10.1021/acsnano.2c10215}

\bibitem[{Mendon{\c{c}}a, Bruno HS and de Moraes, Elizane E and Kirch, Alexsandro and Batista, Ronaldo JC and de Oliveira, Alan B and Barbosa, Marcia C and Chacham, H{\'e}lio(2023)Mendon{\c{c}}a, de~Moraes, Kirch, Batista, de~Oliveira, Barbosa, and Chacham}]{@10.1021/acs.jpcb.3c02889}
Mendon{\c{c}}a BH, de~Moraes EE, Kirch A, Batista RJ, de~Oliveira AB, Barbosa MC, and Chacham H (2023) Flow through deformed carbon nanotubes predicted by rigid and flexible water models. The Journal of Physical Chemistry B 127(40):8634--8643. \doi{10.1021/acs.jpcb.3c02889}

\bibitem[{Mendon{\c{c}}a, Bruno HS and Pereira, Neuma and Rezende, Natália P and Moraes, Elizane E de and Lacerda, Rodrigo G and Chacham, Helio(2024)Mendon{\c{c}}a, Pereira, Rezende, Moraes, Lacerda, and Chacham}]{10.1021/acs.jpcc.4c00078}
Mendon{\c{c}}a BH, Pereira N, Rezende NP, Moraes EEd, Lacerda RG, and Chacham H (2024) Conduction percolation in mos2 nanoink humidity sensors: Critical exponents and nanochannel dimensionality. The Journal of Physical Chemistry C 128(19):8042--8047. \doi{https://doi.org/10.1021/acs.jpcc.4c00078}

\bibitem[{de Moraes, Elizane E and Tonel, Mariana Z and Fagan, Solange B and Barbosa, Marcia C(2019)de~Moraes, Tonel, Fagan, and Barbosa}]{de2019density}
de~Moraes EE, Tonel MZ, Fagan SB, and Barbosa MC (2019) Density functional theory study of $\pi$-aromatic interaction of benzene, phenol, catechol, dopamine isolated dimers and adsorbed on graphene surface. Journal of Molecular Modeling 25(10):302. \doi{https://doi.org/10.1007/s00894-019-4185-2}

\bibitem[{Han, Zhen-yang and Huang, Lin-jun and Qu, Huai-jiao and Wang, Yan-xin and Zhang, Zhi-jie and Rong, Qing-lin and Sang, Zi-qi and Wang, Yao and Kipper, Matt J and Tang, Jian-guo(2021)Han, Huang, Qu, Wang, Zhang, Rong, Sang, Wang, Kipper, and Tang}]{han2021review}
Han Zy, Huang Lj, Qu Hj, Wang Yx, Zhang Zj, Rong Ql, Sang Zq, Wang Y, Kipper MJ, and Tang Jg (2021) A review of performance improvement strategies for graphene oxide-based and graphene-based membranes in water treatment. Journal of Materials Science 56(16):9545--9574. \doi{https://doi.org/10.1007/s10853-021-05873-7}

\bibitem[{Dong, Renhao and Zhang, Tao and Feng, Xinliang(2018)Dong, Zhang, and Feng}]{dong2018interface}
Dong R, Zhang T, and Feng X (2018) Interface-assisted synthesis of 2d materials: trend and challenges. Chemical Reviews 118(13):6189--6235. \doi{https://doi.org/10.1021/acs.chemrev.8b00056}

\bibitem[{Nezval, David and Barto{\v{s}}{\'\i}k, Miroslav and Mach, Jind{\v{r}}ich and {\v{S}}varc, Vojt{\v{e}}ch and Kone{\v{c}}n{\`y}, Martin and Piastek, Jakub and {\v{S}}pa{\v{c}}ek, Ond{\v{r}}ej and {\v{S}}ikola, Tom{\'a}{\v{s}}(2023)Nezval, Barto{\v{s}}{\'\i}k, Mach, {\v{S}}varc, Kone{\v{c}}n{\`y}, Piastek, {\v{S}}pa{\v{c}}ek, and {\v{S}}ikola}]{nezval2023dft}
Nezval D, Barto{\v{s}}{\'\i}k M, Mach J, {\v{S}}varc V, Kone{\v{c}}n{\`y} M, Piastek J, {\v{S}}pa{\v{c}}ek O, and {\v{S}}ikola T (2023) Dft study of water on graphene: Synergistic effect of multilayer p-doping. The Journal of Chemical Physics 159(21). \doi{https://doi.org/10.1063/5.0161160}

\bibitem[{Verma, Ashutosh Kumar and Sharma, Bharat Bhushan(2024)Verma and Sharma}]{verma2024experimental}
Verma AK and Sharma BB (2024) Experimental and theoretical insights into interfacial properties of 2d materials for selective water transport membranes: A critical review. Langmuir 40(15):7812--7834. \doi{https://doi.org/10.1021/acs.langmuir.4c00061}

\bibitem[{Gbadamasi, Sharafadeen and Mohiuddin, Md and Krishnamurthi, Vaishnavi and Verma, Rajni and Khan, Muhammad Waqas and Pathak, Saurabh and Kalantar-Zadeh, Kourosh and Mahmood, Nasir(2021)Gbadamasi, Mohiuddin, Krishnamurthi, Verma, Khan, Pathak, Kalantar-Zadeh, and Mahmood}]{gbadamasi2021interface}
Gbadamasi S, Mohiuddin M, Krishnamurthi V, Verma R, Khan MW, Pathak S, Kalantar-Zadeh K, and Mahmood N (2021) Interface chemistry of two-dimensional heterostructures--fundamentals to applications. Chemical Society Reviews 50(7):4684--4729. \doi{https://doi.org/10.1039/D0CS01070G}

\bibitem[{Bjorneholm, Olle and Hansen, Martin H. and Hodgson, Andrew and Liu, Li-Min and Limmer, David T. and Michaelides, Angelos and Pedevilla, Philipp and Rossmeisl, Jan and Shen, Huaze and Tocci, Gabriele and Tyrode, Eric and Walz, Marie-Madeleine and Werner, Joerg and Bluhm, Hendrik and Pettersson, Lars G. M. and Nilsson, Anders(2016)Bjorneholm, Hansen, Hodgson, Liu, Limmer, Michaelides, Pedevilla, Rossmeisl, Shen, Tocci, Tyrode, Walz, Werner, Bluhm, Pettersson, and Nilsson}]{bjorneholm2016water}
Bjorneholm O, Hansen MH, Hodgson A, Liu LM, Limmer DT, Michaelides A, Pedevilla P, Rossmeisl J, Shen H, Tocci G, Tyrode E, Walz MM, Werner J, Bluhm H, Pettersson LGM, and Nilsson A (2016) Water at interfaces. Chemical Reviews 116(13):7698--7726. \doi{10.1021/acs.chemrev.6b00045}

\bibitem[{Sacchi, Marco and Tamt{\"o}gl, Anton(2023)Sacchi and Tamt{\"o}gl}]{sacchi2023water}
Sacchi M and Tamt{\"o}gl A (2023) Water adsorption and dynamics on graphene and other 2d materials: computational and experimental advances. Advances in Physics: X 8(1):2134051. \doi{https://doi.org/10.1080/23746149.2022.2134051}

\bibitem[{Li, Qiang and Song, Jie and Besenbacher, Flemming and Dong, Mingdong(2015)Li, Song, Besenbacher, and Dong}]{li2015two}
Li Q, Song J, Besenbacher F, and Dong M (2015) Two-dimensional material confined water. Accounts of Chemical Research 48(1):119--127. \doi{https://doi.org/10.1021/ar500306w}

\bibitem[{Kang, Junhyeok and Kwon, Ohchan and Kim, Jeong Pil and Kim, Ju Yeon and Kim, Jiwon and Cho, Yonghwi and Kim, Dae Woo(2024)Kang, Kwon, Kim, Kim, Kim, Cho, and Kim}]{kang2024graphene}
Kang J, Kwon O, Kim JP, Kim JY, Kim J, Cho Y, and Kim DW (2024) Graphene membrane for water-related environmental application: A comprehensive review and perspectives. ACS Environmental Au 5(1):35--60. \doi{https://doi.org/10.1021/acsenvironau.4c00088}

\bibitem[{Hamada, Ikutaro(2012)Hamada}]{hamada2012adsorption}
Hamada I (2012) Adsorption of water on graphene: A van der waals density functional study. Physical Review B—Condensed Matter and Materials Physics 86(19):195436. \doi{https://doi.org/10.1103/PhysRevB.86.195436}

\bibitem[{Peng, Weijun and Li, Hongqiang and Liu, Yanyan and Song, Shaoxian(2017)Peng, Li, Liu, and Song}]{peng2017review}
Peng W, Li H, Liu Y, and Song S (2017) A review on heavy metal ions adsorption from water by graphene oxide and its composites. Journal of Molecular Liquids 230:496--504. \doi{https://doi.org/10.1016/j.molliq.2017.01.064}

\bibitem[{Ma, Jie and Michaelides, Angelos and Alfe, Dario and Schimka, Laurids and Kresse, Georg and Wang, Enge(2011)Ma, Michaelides, Alfe, Schimka, Kresse, and Wang}]{ma2011adsorption}
Ma J, Michaelides A, Alfe D, Schimka L, Kresse G, and Wang E (2011) Adsorption and diffusion of water on graphene from first principles. Physical Review B—Condensed Matter and Materials Physics 84(3):033402. \doi{https://doi.org/10.1103/PhysRevB.84.033402}

\bibitem[{Voloshina, Elena and Usvyat, Denis and Sch{\"u}tz, Martin and Dedkov, Yuriy and Paulus, Beate(2011)Voloshina, Usvyat, Sch{\"u}tz, Dedkov, and Paulus}]{voloshina2011physisorption}
Voloshina E, Usvyat D, Sch{\"u}tz M, Dedkov Y, and Paulus B (2011) On the physisorption of water on graphene: a ccsd (t) study. Physical Chemistry Chemical Physics 13(25):12041--12047. \doi{https://doi.org/10.1039/C1CP20609E}

\bibitem[{Melios, Christos and Giusca, Cristina E and Panchal, Vishal and Kazakova, Olga(2018)Melios, Giusca, Panchal, and Kazakova}]{melios2018water}
Melios C, Giusca CE, Panchal V, and Kazakova O (2018) Water on graphene: review of recent progress. 2D Materials 5(2):022001. \doi{10.1088/2053-1583/aa9ea9}

\bibitem[{Wehling, Tim O and Lichtenstein, Alexander I and Katsnelson, Mikhail I(2008)Wehling, Lichtenstein, and Katsnelson}]{wehling2008first}
Wehling TO, Lichtenstein AI, and Katsnelson MI (2008) First-principles studies of water adsorption on graphene: The role of the substrate. Applied Physics Letters 93(20). \doi{https://doi.org/10.1063/1.3033202}

\bibitem[{Leenaerts, O and Partoens, B and Peeters, FM(2009)Leenaerts, Partoens, and Peeters}]{leenaerts2009water}
Leenaerts O, Partoens B, and Peeters F (2009) Water on graphene: Hydrophobicity and dipole moment using density functional theory. Physical Review B—Condensed Matter and Materials Physics 79(23):235440. \doi{https://doi.org/10.1103/PhysRevB.79.235440}

\bibitem[{Holt, Jason K and Park, Hyung Gyu and Wang, Yinmin and Stadermann, Michael and Artyukhin, Alexander B and Grigoropoulos, Costas P and Noy, Aleksandr and Bakajin, Olgica(2006)Holt, Park, Wang, Stadermann, Artyukhin, Grigoropoulos, Noy, and Bakajin}]{holt2006fast}
Holt JK, Park HG, Wang Y, Stadermann M, Artyukhin AB, Grigoropoulos CP, Noy A, and Bakajin O (2006) Fast mass transport through sub-2-nanometer carbon nanotubes. Science 312(5776):1034--1037. \doi{10.1126/science.1126298}

\bibitem[{Hummer, G and Rasaiah, JC and Noworyta, JP(2002)Hummer, Rasaiah, and Noworyta}]{hummer2002water}
Hummer G, Rasaiah J, and Noworyta J (2002) Water conduction through carbon nanotubes. In: Tech. Proc. 2nd Int. Conf. Comput. Nanosci. Nanotechnol, pp 124--27

\bibitem[{Li, Wen and Wang, Wensen and Zheng, Xin and Dong, Zihan and Yan, Youguo and Zhang, Jun(2017)Li, Wang, Zheng, Dong, Yan, and Zhang}]{li2017molecular}
Li W, Wang W, Zheng X, Dong Z, Yan Y, and Zhang J (2017) Molecular dynamics simulations of water flow enhancement in carbon nanochannels. Computational Materials Science 136:60--66. \doi{https://doi.org/10.1016/j.commatsci.2017.04.024}

\bibitem[{Brandenburg, Jan Gerit and Zen, Andrea and Alf{\`e}, Dario and Michaelides, Angelos(2019)Brandenburg, Zen, Alf{\`e}, and Michaelides}]{brandenburg2019interaction}
Brandenburg JG, Zen A, Alf{\`e} D, and Michaelides A (2019) Interaction between water and carbon nanostructures: How good are current density functional approximations? The Journal of Chemical Physics 151(16). \doi{https://doi.org/10.1063/1.5121370}

\bibitem[{Ahlstr{\"o}m, Peter and Wallqvist, Anders and Engstr{\"o}m, Sven and J{\"o}nsson, Bo(1989)Ahlstr{\"o}m, Wallqvist, Engstr{\"o}m, and J{\"o}nsson}]{ahlstrom1989molecular}
Ahlstr{\"o}m P, Wallqvist A, Engstr{\"o}m S, and J{\"o}nsson B (1989) A molecular dynamics study of polarizable water. Molecular Physics 68(3):563--581. \doi{https://doi.org/10.1080/00268978900102361}

\bibitem[{Lamoureux, Guillaume and Harder, Edward and Vorobyov, Igor V and Roux, Beno{\^\i}t and MacKerell Jr, Alexander D(2006)Lamoureux, Harder, Vorobyov, Roux, and MacKerell~Jr}]{lamoureux2006polarizable}
Lamoureux G, Harder E, Vorobyov IV, Roux B, and MacKerell~Jr AD (2006) A polarizable model of water for molecular dynamics simulations of biomolecules. Chemical Physics Letters 418(1-3):245--249. \doi{https://doi.org/10.1016/j.cplett.2005.10.135}

\bibitem[{Cao, Chao and Wu, Min and Jiang, Jianzhong and Cheng, Hai-Ping(2010)Cao, Wu, Jiang, and Cheng}]{cao2010transition}
Cao C, Wu M, Jiang J, and Cheng HP (2010) Transition metal adatom and dimer adsorbed on graphene: Induced magnetization and electronic structures. Physical Review B—Condensed Matter and Materials Physics 81(20):205424. \doi{https://doi.org/10.1103/PhysRevB.81.205424}

\bibitem[{Vanin, Marco and Gath, Jesper and Thygesen, Kristian Sommer and Jacobsen, Karsten Wedel(2010)Vanin, Gath, Thygesen, and Jacobsen}]{vanin2010first}
Vanin M, Gath J, Thygesen KS, and Jacobsen KW (2010) First-principles calculations of graphene nanoribbons in gaseous environments: Structural and electronic properties. Physical Review B—Condensed Matter and Materials Physics 82(19):195411. \doi{https://doi.org/10.1103/PhysRevB.82.195411}

\bibitem[{Michaelides, Angelos(2006)Michaelides}]{michaelides2006density}
Michaelides A (2006) Density functional theory simulations of water--metal interfaces: waltzing waters, a novel 2d ice phase, and more. Applied Physics A 85(4):415--425. \doi{https://doi.org/10.1007/s00339-006-3695-9}

\bibitem[{Valle, Jo{\~a}o VL and Mendon{\c{c}}a, Bruno HS and Barbosa, Marcia C and Chacham, Helio and de Moraes, Elizane E(2024)Valle, Mendon{\c{c}}a, Barbosa, Chacham, and de~Moraes}]{valle2024accuracy}
Valle JV, Mendon{\c{c}}a BH, Barbosa MC, Chacham H, and de~Moraes EE (2024) Accuracy of tip4p/2005 and spc/fw water models. The Journal of Physical Chemistry B 128(4):1091--1097. \doi{https://doi.org/10.1021/acs.jpcb.3c07044}

\bibitem[{Falk, Kerstin and Sedlmeier, Felix and Joly, Laurent and Netz, Roland R and Bocquet, Lyd{\'e}ric(2010)Falk, Sedlmeier, Joly, Netz, and Bocquet}]{falk2010molecular}
Falk K, Sedlmeier F, Joly L, Netz RR, and Bocquet L (2010) Molecular origin of fast water transport in carbon nanotube membranes: superlubricity versus curvature dependent friction. Nano Letters 10(10):4067--4073. \doi{https://doi.org/10.1021/nl1021046}

\bibitem[{Singh, Harpreet and Myong, Rho Shin(2018)Singh and Myong}]{singh2018critical}
Singh H and Myong RS (2018) Critical review of fluid flow physics at micro-to nano-scale porous media applications in the energy sector. Advances in Materials Science and Engineering 2018(1):9565240. \doi{https://doi.org/10.1155/2018/9565240}

\bibitem[{Shuvo, Abdul Aziz and Paniagua-Guerra, Luis E and Choi, Juseok and Kim, Seong H and Ramos-Alvarado, Bladimir(2025)Shuvo, Paniagua-Guerra, Choi, Kim, and Ramos-Alvarado}]{shuvo2025hydrodynamic}
Shuvo AA, Paniagua-Guerra LE, Choi J, Kim SH, and Ramos-Alvarado B (2025) Hydrodynamic slip in nanoconfined flows: a review of experimental, computational, and theoretical progress. Nanoscale 17(2):635--660. \doi{10.1039/D4NR03697B}

\bibitem[{Xu, Zhi and Wu, Nan and Abdelghani-Idrissi, Soufiane and Tr{\'e}gou{\"e}t, Corentin and Perez-Carvajal, Javier and Colin, Annie and Ma, Ming and Nigu{\`e}s, Antoine and Siria, Alessandro(2025)Xu, Wu, Abdelghani-Idrissi, Tr{\'e}gou{\"e}t, Perez-Carvajal, Colin, Ma, Nigu{\`e}s, and Siria}]{xu2025advanced}
Xu Z, Wu N, Abdelghani-Idrissi S, Tr{\'e}gou{\"e}t C, Perez-Carvajal J, Colin A, Ma M, Nigu{\`e}s A, and Siria A (2025) Advanced nanoscale functionalities for water and energy technologies. Advanced Physics Research p 2400195. \doi{https://doi.org/10.1002/apxr.202400195}

\bibitem[{Cohen-Tanugi, David and Grossman, Jeffrey C(2012)Cohen-Tanugi and Grossman}]{cohen2012water}
Cohen-Tanugi D and Grossman JC (2012) Water desalination across nanoporous graphene. Nano Letters 12(7):3602--3608. \doi{https://doi.org/10.1021/nl3012853}

\bibitem[{Surwade, Sumedh P and Smirnov, Sergei N and Vlassiouk, Ivan V and Unocic, Raymond R and Veith, Gabriel M and Dai, Sheng and Mahurin, Shannon M(2015)Surwade, Smirnov, Vlassiouk, Unocic, Veith, Dai, and Mahurin}]{surwade2015water}
Surwade SP, Smirnov SN, Vlassiouk IV, Unocic RR, Veith GM, Dai S, and Mahurin SM (2015) Water desalination using nanoporous single-layer graphene. Nature Nanotechnology 10(5):459--464. \doi{https://doi.org/10.1038/nnano.2015.37}

\bibitem[{Homaeigohar, Shahin and Elbahri, Mady(2017)Homaeigohar and Elbahri}]{homaeigohar2017graphene}
Homaeigohar S and Elbahri M (2017) Graphene membranes for water desalination. NPG Asia Materials 9(8):e427--e427. \doi{https://doi.org/10.1038/am.2017.135}

\bibitem[{Konatham, Deepthi and Yu, Jing and Ho, Tuan A and Striolo, Alberto(2013)Konatham, Yu, Ho, and Striolo}]{konatham2013simulation}
Konatham D, Yu J, Ho TA, and Striolo A (2013) Simulation insights for graphene-based water desalination membranes. Langmuir 29(38):11884--11897. \doi{https://doi.org/10.1021/la4018695}

\bibitem[{Aghigh, Arash and Alizadeh, Vahid and Wong, Hin Yong and Islam, Md Shabiul and Amin, Nowshad and Zaman, Mukter(2015)Aghigh, Alizadeh, Wong, Islam, Amin, and Zaman}]{aghigh2015recent}
Aghigh A, Alizadeh V, Wong HY, Islam MS, Amin N, and Zaman M (2015) Recent advances in utilization of graphene for filtration and desalination of water: A review. Desalination 365:389--397. \doi{https://doi.org/10.1016/j.desal.2015.03.024}

\bibitem[{Kohn, W. and Sham, L. J.(1965)Kohn and Sham}]{PhysRev.140.A1133}
Kohn W and Sham LJ (1965) Self-consistent equations including exchange and correlation effects. Phys Rev 140:A1133--A1138. \doi{10.1103/PhysRev.140.A1133}

\bibitem[{José M. Soler and Emilio Artacho and Julian D. Gale and Alberto García and Javier Junquera and Pablo Ordejón and Daniel Sánchez-Portal(2002)Soler, Artacho, Gale, García, Junquera, Ordejón, and Sánchez-Portal}]{soler2002siesta}
Soler JM, Artacho E, Gale JD, García A, Junquera J, Ordejón P, and Sánchez-Portal D (2002) The siesta method for ab initio order- n materials simulation. J Phys: Condens Matter 14(11):2745. \doi{10.1088/0953-8984/14/11/302}

\bibitem[{S{\'a}nchez-Portal, Daniel and Ordejon, Pablo and Artacho, Emilio and Soler, Jose M(1997)S{\'a}nchez-Portal, Ordejon, Artacho, and Soler}]{sanchez1997density}
S{\'a}nchez-Portal D, Ordejon P, Artacho E, and Soler JM (1997) Density-functional method for very large systems with lcao basis sets. Int J Quantum Chem 65(5):453--461. \doi{https://doi.org/10.1002/(SICI)1097-461X(1997)65:5<453::AID-QUA9>3.0.CO;2-V}

\bibitem[{Cooper, Valentino R(2010)Cooper}]{cooper2010van}
Cooper VR (2010) Van der waals density functional: An appropriate exchange functional. Phys Rev B 81(16):161104. \doi{10.1103/PhysRevB.81.161104}

\bibitem[{L. Kleinman and D. M. Bylander(1982)Kleinman and Bylander}]{testeKB}
Kleinman L and Bylander DM (1982) Efficacious form for model pseudopotentials. Phys Rev Lett 48:1425--1428. \doi{10.1103/PhysRevLett.48.1425}

\bibitem[{Troullier, N. and Martins, J. L.(1991)Troullier and Martins}]{martins91}
Troullier N and Martins JL (1991) Efficient pseudopotentials for plane-wave calculations. Phys Rev B 43:1993--2006. \doi{10.1103/PhysRevLett.48.1425}

\bibitem[{Raj, R Ranga and Sathish, S and Mansadevi, TLD and Supriya, R and Sekar, S and Patil, Pravin P and Tonmoy, Mahtab Mashuq(2022)Raj, Sathish, Mansadevi, Supriya, Sekar, Patil, and Tonmoy}]{raj2022effect}
Raj RR, Sathish S, Mansadevi T, Supriya R, Sekar S, Patil PP, and Tonmoy MM (2022) Effect of graphene fillers on the water absorption and mechanical properties of naoh-treated kenaf fiber-reinforced epoxy composites. Journal of Nanomaterials 2022(1):1748121. \doi{https://doi.org/10.1155/2022/1748121}

\bibitem[{Carlson, Shane R and Schullian, Otto and Becker, Maximilian R and Netz, Roland R(2024)Carlson, Schullian, Becker, and Netz}]{carlson2024modeling}
Carlson SR, Schullian O, Becker MR, and Netz RR (2024) Modeling water interactions with graphene and graphite via force fields consistent with experimental contact angles. The Journal of Physical Chemistry Letters 15(24):6325--6333. \doi{https://doi.org/10.1021/acs.jpclett.4c01143}

\bibitem[{Pizzochero, Michele and Tepliakov, Nikita V and Lischner, Johannes and Mostofi, Arash A and Kaxiras, Efthimios(2024)Pizzochero, Tepliakov, Lischner, Mostofi, and Kaxiras}]{pizzochero2024one}
Pizzochero M, Tepliakov NV, Lischner J, Mostofi AA, and Kaxiras E (2024) One-dimensional magnetic conduction channels across zigzag graphene nanoribbon/hexagonal boron nitride heterojunctions. Nano Letters 24(22):6521--6528. \doi{https://doi.org/10.1021/acs.nanolett.4c00920}

\bibitem[{Temmen, M and Ochedowski, O and Schleberger, M and Reichling, M and Bollmann, TRJ(2014)Temmen, Ochedowski, Schleberger, Reichling, and Bollmann}]{temmen2014hydration}
Temmen M, Ochedowski O, Schleberger M, Reichling M, and Bollmann T (2014) Hydration layers trapped between graphene and a hydrophilic substrate. New Journal of Physics 16(5):053039. \doi{10.1088/1367-2630/16/5/053039}

\end{thebibliography}

\end{document}